\documentclass[prd,twocolumn,showpacs,floatfix,amsmath,nofootinbib,amssymb,floatfix]{revtex4}
\usepackage{graphicx,color,dcolumn,booktabs,bm}
\usepackage{longtable,lscape}
\usepackage{txfonts}
\usepackage{overpic}
\usepackage{amssymb}
\usepackage{indentfirst}
\usepackage{feynmf}   
\usepackage{slashed}  
\usepackage{cases}
\usepackage{color}
\usepackage{multirow}
\usepackage{epstopdf}
\usepackage{longtable}
\usepackage{graphicx,color,dcolumn,booktabs,bm}
\usepackage[colorlinks,
            citecolor=blue,
            anchorcolor=red,
            menucolor=red,
            linkcolor=red,
            filecolor=red,
            runcolor=red,
            urlcolor=blue,
            frenchlinks=red]{hyperref}

\graphicspath{{Figures/}} %

\allowdisplaybreaks

\begin{document}

\title{Charmoniumlike resonant explanation on the newly observed $X(3960)$}

\author{Rui Chen$^{1}$}\email{chenrui@hunnu.edu.cn}
\author{Qi Huang$^{2}$\footnote{Corresponding author}}\email{huangqi@ucas.ac.cn}

\affiliation{
$^1$Key Laboratory of Low-Dimensional Quantum Structures and Quantum Control of Ministry of Education, Department of Physics and Synergetic Innovation Center for Quantum Effects and Applications, Hunan Normal University, Changsha 410081, China\\
$^2$Department of Physics, Nanjing Normal University, Nanjing 210023, China}
\date{\today}

\begin{abstract}
Stimulated by the observation of the newly $X(3960)$ observed by the LHCb collaboration, we adopt the one-boson-exchange model and consider the $S-D$ wave mixing effects to study the $D_s\bar{D}_s/D^*\bar{D}^*/D_s^*\bar{D}_s^*$ interactions with $I(J^{PC})=0(0^{++})$. After producing the phase shifts of this coupled channel systems, our results show that there can exist a charmoniumlike resonance, whose obtained mass and width can both well match with the experimental data of the newly observed $X(3960)$. We also find that the $D^*\bar{D}^*$ system plays an important role in the formation of the newly observed $X(3960)$ as a charmoniumlike resonance, and the $D_s^*\bar{D}_s^*$ system makes a significant contribution to the resonant width. As a byproduct, we perform a coupled channel analysis on the $D^*\bar{D}^*/D_s\bar{D}_s^*/D_s^*\bar{D}_s^*$ interactions with $I(J^{PC})=0(1^{+-})$, our results can predict the existence of the $D_s\bar{D}_s^*$ molecule with $1^{+-}$ and the $D_s^*\bar{D}_s^*$ molecule with $1^{+-}$. Their widths are around several and several to several tens MeV, respectively. Experimental searches for these two possible charmoniumlike molecular candidates can be helpful to verify our proposal.
\end{abstract}

\pacs{12.39.Pn, 13.75.Lb, 14.40.Rt}

\maketitle

\section{introduction}

Recently, in a talk given at CERN, the LHCb collaboration reported their observations of three new states from the $B$ decay processes \cite{LHCb:Tcc}. In this talk, apart from the two charged states, i.e., the first pentaquark with $s$ quark content observed in the $J/\psi \Lambda$ invariant mass spectrum of the $B^- \to J/\psi \Lambda \bar{p}$ process and the $T_{c\bar{s}}^{a++(0)}(2900)$ shown in the $D_s^+ \pi^{+/(-)}$ invariant mass spectrum of the $B^{0(+)} \to D^{0(-)} D_s^+ \pi^{+(-)}$ process, one another neutral state, namely $X(3960)$, was also observed by the LHCb Collaboration in the $D_s^+ D_s^-$ invariant mass spectrum of the $B$ decay process $B^+ \to D_s^+ D_s^- K^+$ \cite{LHCb:Tcc}.

Usually, when a new neutral state is observed in the invariant mass spectrum that composed by a pair of heavy and anti-heavy mesons, our first consideration may often be that if this new state can be treated as a conventional charmonia. Here, since the quantum number of the $X(3960)$ is reported as $0^{++}$, the first idea of us is that if it is a new $\chi_{c0}$ state. When checking the theoretical results of the potential model \cite{Barnes:2005pb,Li:2009zu,Duan:2020tsx}, the $\chi_{c0}(2P)$ state has been denoted as $\chi_{c0}(3915)$ \cite{Workman:2022ynf} and the position of $\chi_{c0}(3P)$ is around 4.2 GeV. For $\chi_{c0}(3915)$, although its mass is close to the $X(3960)$, whose mass and width are measured as $M= 3955 \pm 6 \pm 22$ MeV and $\Gamma = 48 \pm 17 \pm 10$ MeV respectively, due to its mass is below the threshold of $D_s \bar{D}_s$, i.e., about 3938 MeV, it is puzzled that it can be observed in the $D_s \bar{D}_s$ invariant mass spectrum. While for the $\chi_{c0}(3P)$, its predicted mass is too far away from $X(3960)$, thus denoting the $X(3960)$ as $\chi_{c0}(3P)$ may be not appropriate.

Another reason that $X(3960)$ may not be a good candidate of charmonia is that its decay property is a little different. In the talk \cite{LHCb:Tcc}, the LHCb Collaboration compared its decay widths to the $D^+ D^-$ and $D_s^+ D_s^-$, and the measurement gave that \cite{LHCb:Tcc}
\begin{eqnarray}
	\frac{\Gamma(X(3960) \to D^+ D^-)}{\Gamma(X(3960) \to D_s^+ D_s^-)}=0.29 \pm 0.09 \pm 0.10 \pm 0.08,
\end{eqnarray}
which means it is easier for $X(3960)$ to decay into $D_s^+ D_s^-$ rather than $D^+ D^-$. Since usually it is harder to excite a pair of $s\bar{s}$ from vacuum compared with $u\bar{u}(d\bar{d})$, thus conventional charmonia predominantly decay into a pair of $D$ meson, which implies the exotic nature of this new state $X(3960)$ \cite{LHCb:Tcc}.

Thus, the next thing we do is naturally to see if this $X(3960)$ can really be assigned as an exotic state, in which a state composed by four valence quarks may be the easist generalization. Since the position of $X(3960)$ is close to the $D_s \bar{D}_s$ threshold, a consideration that it is related to some molecular states can easily raise up. Actually, studies on the $D_s \bar{D}_s$ molecular states has already done before the observation of $X(3960)$ \cite{Prelovsek:2020eiw,Gamermann:2006nm,Nieves:2012tt,Hidalgo-Duque:2012rqv,Meng:2020cbk,Dong:2021juy}, and it turns out that although a $0^{++}$ bound state that couples strongly to $D_s^+D_s^-$ and weakly to $D^+ D^-$ is found just below $D_s^+ D_s^-$ threshold, after carrying out a dynamic study of $D\bar{D}$ and $D_s^+ D_s^-$ in coupled channels, this bound state disappears. Thus, recently, Ref. \cite{Bayar:2022dqa} reanalyzed this situation and found that if the strength of $D\bar{D} \to D_s^+ D_s^-$ transition is slightly reduced, that missing state will appear and behave similarly on the $D_s^+ D_s^-$ invariant mass spectrum as the experimental observation \cite{LHCb:Tcc,Bayar:2022dqa}.

The molecular state interpretion of the $X(3960)$ is also supported by Refs. \cite{Xin:2022bzt,Ji:2022uie}. In addition, apart from bound state interpretion, Ref. \cite{Ji:2022uie} pointed out that virtual state explanation was also valid. Then, Ref. \cite{Xie:2022lyw} used the effective Lagrangian approach to calculate the production rate of $X(3960)$ in the $B$ decays utilizing triangle diagrams, and the results showed that both the bound and virtual state interpretions can match the relevant experimental data.

Thus, for $X(3960)$ appeared in the $D_s^+ D_s^-$ invariant mass spectrum, Refs. \cite{Prelovsek:2020eiw,Gamermann:2006nm,Nieves:2012tt,Hidalgo-Duque:2012rqv,Meng:2020cbk,Dong:2021juy,Bayar:2022dqa,Xin:2022bzt,Ji:2022uie,Xie:2022lyw} explain it as an effect caused by a molecular state located below the $D_s^+ D_s^-$ threshold. However, considering the fact that its measured mass is above the $D_s^+ D_s^-$ threshold \cite{LHCb:Tcc}, the resonant state explanation, in our view, is also possible, and this work is to study this possibility.

In general, the resonances can be divided into two types, i.e., the shape-type resonances and the Feshbach-type resonances, and the generation of these two types of resonances is controlled by the potential barriers \cite{Chen:2020yvq}. In addition, we want to emphasize here that the coupled channel effect plays a very important role in producing the Feshbach-type resonances since the mass gaps between the relevant channels will give additional contributions to the potential barriers.

Thus, in this work, considering the measured mass and quantum number of $X(3960)$, we perform an analysis that includes coupled channel effect and $S-D$ wave mixing effect to see if the newly observed $X(3960)$ can be interpreted as a resonance, the included channels are $D_s \bar{D}_s$, $D^\ast \bar{D}^\ast$, and $D_s^\ast \bar{D}_s^\ast$.

This paper is organized as follows. After this introduction, we deduce the coupled $D_s\bar{D}_s/D^*\bar{D}^*/D_s^*\bar{D}_s^*$ interactions with $I(J^{PC})=0(0^{++})$ by using the OBE model in Sec.~\ref{sec2}. In Sec.~\ref{sec3}, we present the corresponding numerical results by producing the phase shifts and predict possible charmoniumlike structures from the isoscalar $D^*\bar{D}^*/D_s\bar{D}_s^*/D_s^*\bar{D}_s^*$ interactions with $J^{PC}=1^{+-}$. The paper ends with a summary in Sec. \ref{sec4}.

\section{Interactions}\label{sec2}

In the OBE model, the relevant effective potentials can be deduced as follows. Firstly, we can write down the scattering amplitude by adopting the effective Lagrangian approach. And then, one can derive the effective potentials based on the approximation relation to the scattering amplitude,
\begin{eqnarray}\label{breit}
\mathcal{V}_{E}^{AB\to CD}(\bm{q}) &=&
          -\frac{\mathcal{M}(AB\to CD)}
          {\sqrt{\prod_i2M_i\prod_f2M_f}},
\end{eqnarray}
where $\mathcal{M}(AB\to CD)$ denotes the scattering amplitude for the $AB\to CD$ process in $t-$channel. $M_i$ and $M_f$ are the masses of the initial states and final states, respectively. Then we can finally obtain the effective potentials in the coordinate space $\mathcal{V}(\bm{r})$ by performing the Fourier transformation, i.e.,
\begin{eqnarray}
\mathcal{V}_{E}(\bm{r}) =
          \int\frac{d^3\bm{q}}{(2\pi)^3}e^{i\bm{q}\cdot\bm{r}}
          \mathcal{V}_{E}(\bm{q})\mathcal{F}^2(q^2,m_E^2).
\end{eqnarray}
Here, $\mathcal{F}(q^2,m_E^2)$ is the form factor, it is introduced at every interactive vertex to compensate the off-shell effect of the exchanged meson. In this work, we take the monopole type form factor, $\mathcal{F}(q^2,m_E^2)= (\Lambda^2-m_E^2)/(\Lambda^2-q^2)$, where $\Lambda$, $m_E$ and $q$ are the cutoff, the mass and four-momentum of the exchanged meson, respectively.

Based on the heavy quark symmetry and chiral symmetry \cite{Yan:1992gz,Wise:1992hn,Burdman:1992gh,Casalbuoni:1996pg,Falk:1992cx}, the relevant effective Lagrangians are constructed as
\begin{eqnarray}\label{eq:compactlag}
{\mathcal L}&=&g_{\sigma}\left\langle H^{(Q)}_a\sigma\overline{H}^{(Q)}_a\right\rangle+g_{\sigma}\left\langle \overline{H}^{(\bar{Q})}_a\sigma H^{(\bar{Q})}_a\right\rangle\nonumber\\
&&+ig\left\langle H^{(Q)}_b{\mathcal A}\!\!\!\slash_{ba}\gamma_5\overline{H}^{\,({Q})}_a\right\rangle+ig\left\langle \overline{H}^{(\bar{Q})}_a{\mathcal A}\!\!\!\slash_{ab}\gamma_5 H^{\,(\bar{Q})}_b\right\rangle\nonumber\\
&&+\left\langle iH^{(Q)}_b\left(\beta v^{\mu}({\mathcal V}_{\mu}-\rho_{\mu})+\lambda \sigma^{\mu\nu}F_{\mu\nu}(\rho)\right)_{ba}\overline{H}^{\,(Q)}_a\right\rangle\nonumber\\
&&-\left\langle i\overline{H}^{(\bar{Q})}_a\left(\beta v^{\mu}({\mathcal V}_{\mu}-\rho_{\mu})-\lambda \sigma^{\mu\nu}F_{\mu\nu}(\rho)\right)_{ab}H^{\,(\bar{Q})}_b\right\rangle.\label{lag}
\end{eqnarray}
Here, the super-fields $H^{(Q)}_a=(1+{v}\!\!\!\slash)(\mathcal{P}^{*(Q)\mu}_a\gamma_{\mu}-\mathcal{P}^{(Q)}_a\gamma_5)/2$ and $H^{(\overline{Q})}_a = (\bar{\mathcal{P}}^{*(\overline{Q})\mu}_a\gamma_{\mu}-\bar{\mathcal{P}}^{(\overline{Q})}_a\gamma_5)
(1-{v}\!\!\!\slash)/{2}$ are expressed as the combinations of the $S-$wave charmed (anti-charmed) mesons with $J^P=0^-$ and $1^-$ as they are in the same doublet in the heavy quark limit. The conjugate field reads as $\overline{H}=\gamma_0H^{\dagger}\gamma_0$. $\mathcal{P}^{{Q}(*)}$ stands for the pseudoscalar (or vector) mesons fields $\mathcal{P}^{{Q}(*)}=(D^{(*)+},~D^{(*)0},~D_s^{(*)+})^T$. $v^{\mu}$ is the four velocity. In the non-relativistic approximation, we take the form of $v^{\mu}=(1, \bf{0})$. $\mathcal{V} = \frac{1}{2}(\xi^{\dag}\partial_{\mu}\xi+\xi\partial_{\mu}\xi^{\dag})$ and $A_{\mu} = \frac{1}{2}(\xi^{\dag}\partial_{\mu}\xi-\xi\partial_{\mu}\xi^{\dag})$ respectively, which stand for the vector and axial currents with $\xi=\text{exp}(i{\mathbb{P}}/f_{\pi})$, $f_{\pi}=132$ MeV. And
$F_{\mu\nu}(\rho)=\partial_{\mu}\rho_{\nu}-\partial_{\nu}\rho_{\mu}+[\rho_{\mu},\rho_{\nu}]$, with $\rho_{\mu}=ig_{V}{\mathcal{V}}_{\mu}/\sqrt{2}$. $g_V=m_{\rho}/f_{\pi}=5.8$. The $\mathbb{P}$ and $\mathbb{V}$ denote the light pseudoscalar meson and the light vector meson matrices, respectively, which read as
\begin{eqnarray}
\left.\begin{array}{c} {\mathbb{P}} = {\left(\begin{array}{ccc}
       \frac{\pi^0}{\sqrt{2}}+\frac{\eta}{\sqrt{6}} &\pi^+ &K^+\\
       \pi^-       &-\frac{\pi^0}{\sqrt{2}}+\frac{\eta}{\sqrt{6}} &K^0\\
       K^-         &\bar K^0   &-\sqrt{\frac{2}{3}} \eta     \end{array}\right)},\\
{\mathbb{V}} = {\left(\begin{array}{ccc}
       \frac{\rho^0}{\sqrt{2}}+\frac{\omega}{\sqrt{2}} &\rho^+ &K^{*+}\\
       \rho^-       &-\frac{\rho^0}{\sqrt{2}}+\frac{\omega}{\sqrt{2}} &K^{*0}\\
       K^{*-}         &\bar K^{*0}   & \phi     \end{array}\right)}.
\end{array}\right.
\end{eqnarray}

Once expanding the effective Lagrangians in Eq. (\ref{lag}), we can obtain
\begin{eqnarray}
\mathcal{L}_{\mathcal{P}^*\mathcal{P}^*\sigma} &=& 2g_{s}\mathcal{P}_a^*\cdot \mathcal{P}_a^{*\dag}\sigma
            +2g_{s}\bar{\mathcal{P}}_a^{*\dag}\cdot\bar{\mathcal{P}}_a^{*}\sigma,\\
\mathcal{L}_{\mathcal{P}^*\mathcal{P}^*\mathbb{P}} &=&
           -i\frac{2g}{f_{\pi}}v^{\beta}\varepsilon_{\beta\mu\alpha\nu}
           \mathcal{P}_{b}^{*\mu}\mathcal{P}_{a}^{*\nu\dag}\partial^{\alpha}\mathbb{P}_{ba}\nonumber\\
           &&+i\frac{2g}{f_{\pi}}v^{\beta}\varepsilon_{\beta\mu\alpha\nu}
           \bar{\mathcal{P}}_{a}^{*\mu\dag}\bar{\mathcal{P}}_{b}^{*\nu}\partial^{\alpha}\mathbb{P}_{ab},\\
\mathcal{L}_{\mathcal{P}^*\mathcal{P}^*\mathbb{V}} &=& \sqrt{2}\beta g_{V}
            \left(\mathcal{P}_{b}^{*}\cdot \mathcal{P}_{a}^{*\dag}\right)(v\cdot\mathbb{V}_{ba})\nonumber\\
            &&-i2\sqrt{2}\lambda g_{V}\mathcal{P}_{b}^{*\mu}\mathcal{P}_{a}^{*\nu\dag}
            \left(\partial_{\mu}\mathbb{V}_{\nu}-\partial_{\nu}\mathbb{V}_{\mu}\right)_{ba}\nonumber\\
            &&-\sqrt{2}\beta g_{V}
            \left(\bar{\mathcal{P}}_{a}^{*\dag}\cdot\bar{\mathcal{P}}_{b}^{*}\right)(v\cdot\mathbb{V}_{ab})\nonumber\\
            &&-i2\sqrt{2}\lambda g_{V}\bar{\mathcal{P}}_{a}^{*\mu\dag}\bar{\mathcal{P}}_{b}^{*\nu}
            \left(\partial_{\mu}\mathbb{V}_{\nu}-\partial_{\nu}\mathbb{V}_{\mu}\right)_{ab}.
\end{eqnarray}
In the above Lagrangians, the $\sigma$ meson coupling $g_s=2.82$ is estimated from the quark model \cite{Machleidt:1987hj,Wang:2019aoc}. For the $\pi-$exchange coupling, $g=0.59$ is extracted from the decay width of $D^*\to D\pi$ \cite{Casalbuoni:1996pg}. Using the vector meson dominance \cite{Isola:2003fh}, $\beta$ is fixed as $\beta=$0.9. $\lambda=$ 0.56 GeV$^{-1}$ is determined through a comparison of the form factor between the theoretical result and the lattice QCD.

The flavor wave functions $|I=0,I_3=0\rangle$ for the isoscalar $D_s^{(*)}\bar{D}_s^{(*)}$ and $D^*\bar{D}^*$ are constructed as $|D_s^{(*)+}{D}_s^{(*)-}\rangle$ and $(|D^{*0}\bar{D}^{*0}\rangle+|D^{*+}D^{*-}\rangle)/\sqrt{2}$, respectively. When we consider the $S$-$D$ wave mixing effects, the spin-orbit wave functions for the $D_{(s)}^*\bar{D}_{(s)}^*$ systems with $0^{++}$ are $|{}^1\mathbb{S}_{0}\rangle$ and $|{}^5\mathbb{D}_{0}\rangle$. In the isoscalar $D_s\bar{D}_s/D^*\bar{D}^*/D_s^*\bar{D}_s^*$ coupled channel analysis, the OBE effective potentials can be expresses as
\begin{eqnarray}
V &=& \left(\begin{array}{cccc} V^{D_s\bar{D}_s\to D_s\bar{D}_s}
        &V^{D^*\bar{D}^*\to D_s\bar{D}_s}      &V^{D_s^*\bar{D}_s^*\to D_s\bar{D}_s}\\
    V^{D_s\bar{D}_s\to D^*\bar{D}^*}
        &V^{D^*\bar{D}^*\to D^*\bar{D}^*}      &V^{D_s^*\bar{D}_s^*\to D^*\bar{D}^*}\\
    V^{D_s\bar{D}_s\to D_s^*\bar{D}_s^*}
        &V^{D_s\bar{D}_s\to D^*\bar{D}^*}      &V^{D_s^*\bar{D}_s^*\to D_s^*\bar{D}_s^*}\end{array}\right).\label{total}
\end{eqnarray}
The corresponding subpotentials read as
\begin{eqnarray}
V^{D_s\bar{D}_s\to D_s\bar{D}_s} &=& -g_s^2Y(\Lambda,m_{\sigma},r)-\frac{1}{2}\beta^2g_V^2Y(\Lambda,m_{\phi},r),\label{potent1}\\
V^{D_s\bar{D}_s\to D^*\bar{D}^*} &=&
   -\frac{\sqrt{2}}{3}\frac{g^2}{f_{\pi}^2}\mathcal{Z}^{12}_{\Lambda,m_K}
   -\frac{2\sqrt{2}}{3}\lambda^2g_V^2\mathcal{Z}^{\prime 12}_{\Lambda, m_{K^*}},\label{eqdx}\\
V^{D_s\bar{D}_s\to D_s^*\bar{D}_s^*} &=&
   \frac{2}{9}\frac{g^2}{f_{\pi}^2}\mathcal{Z}^{13}_{\Lambda,m_{\eta}}
   -\frac{2}{3}\lambda^2g_V^2\mathcal{Z}^{\prime 13}_{\Lambda,m_{\phi}},\label{eqdsx}\\
V^{D^*\bar{D}^*\to D^*\bar{D}^*} &=& -g_s^2\mathcal{Y}^{22}_{\Lambda,m_{\sigma}}
   -\frac{1}{2}\frac{g^2}{f_{\pi}^2}\mathcal{Z}^{22}_{\Lambda,m_{\pi}}
   -\frac{1}{18}\frac{g^2}{f_{\pi}^2}\mathcal{Z}^{22}_{\Lambda,m_{\eta}}\nonumber\\
   &&-\frac{3}{4}\beta^2g_V^2\mathcal{Y}^{22}_{\Lambda,m_{\rho}}
   +\lambda^2g_V^2\mathcal{Z}^{\prime 22}_{\Lambda,m_{\rho}}\nonumber\\
   &&-\beta^2g_V^2\mathcal{Y}^{22}_{\Lambda,m_{\omega}}
   +\frac{1}{3}\lambda^2g_V^2\mathcal{Z}^{\prime 22}_{\Lambda,m_{\omega}},\\
V^{D^*\bar{D}^*\to D_s^*\bar{D}_s^*} &=&
   -\frac{\sqrt{2}}{3}\frac{g^2}{f_{\pi}^2}\mathcal{Z}^{13}_{\Lambda,m_K}
   -\frac{\beta^2g_V^2}{\sqrt{2}}\mathcal{Y}^{13}_{\Lambda,m_{K^*}}\nonumber\\
   &&+\frac{2\sqrt{2}}{3}\lambda^2g_V^2\mathcal{Z}^{\prime 13}_{\Lambda, m_{K^*}},\\
V^{D_s^*\bar{D}_s^*\to D_s^*\bar{D}_s^*} &=& -g_s^2\mathcal{Y}^{33}_{\Lambda,m_{\sigma}}
   -\frac{2}{9}\frac{g^2}{f_{\pi}^2}\mathcal{Z}^{33}_{\Lambda,m_{\eta}}
   -\frac{\beta^2g_V^2}{2}\mathcal{Y}^{33}_{\Lambda,m_{\phi}}\nonumber\\
   &&+\frac{2}{3}\lambda^2g_V^2\mathcal{Z}^{\prime 33}_{\Lambda, m_{\phi}}.\label{potent2}
\end{eqnarray}
Here, we define several useful functions as follows, i.e.,
\begin{eqnarray}
Y(\Lambda,m,{r}) &=& \frac{1}{4\pi r}(e^{-mr}-e^{-\Lambda
r})-\frac{\Lambda^2-m^2}{8\pi \Lambda}e^{-\Lambda r},\\
\mathcal{Y}^{ij}_{\Lambda, m_a}&=&\mathcal{D}_{ij}Y(\Lambda,m_a,r),\\
\mathcal{Z}^{ij}_{\Lambda, m_a}&=&\left(\mathcal{E}_{ij}\nabla^2+\mathcal{F}_{ij}r\frac{\partial}{\partial r}\frac{1}{r}\frac{\partial}{\partial r}\right)Y(\Lambda,m_a,r),\\
\mathcal{Z}^{\prime ij}_{\Lambda, m_a} &=&\left(2\mathcal{E}_{ij}\nabla^2-\mathcal{F}_{ij}r\frac{\partial}{\partial
r}\frac{1}{r}\frac{\partial}{\partial r}\right)Y(\Lambda,m_a,r).
\end{eqnarray}
In above effective potentials (\ref{potent1})-(\ref{potent2}), $\mathcal{D}_{ij}$, $\mathcal{E}_{ij}$, and
$\mathcal{F}_{ij}$ stand for the operators for the spin-spin interactions and the tensor forces, respectively. For example, $\mathcal{E}_{12}=\mathcal{E}_{13}
=\bm{\epsilon}_3^{\dag}\cdot\bm{\epsilon}_4^{\dag}$, $\mathcal{F}_{12}=\mathcal{F}_{13}
=S(\hat{r},\bm{\epsilon}_3^{\dag},\bm{\epsilon}_4^{\dag})$, $\mathcal{D}_{22}=\mathcal{D}_{33}=
(\bm{\epsilon}_1\cdot\bm{\epsilon}_3^{\dag})(\bm{\epsilon}_2\cdot\bm{\epsilon}_4^{\dag})$, $\mathcal{E}_{22}=\mathcal{E}_{23}=\mathcal{E}_{33}=
(\bm{\epsilon}_1\times\bm{\epsilon}_3^{\dag})\cdot(\bm{\epsilon}_2\times\bm{\epsilon}_4^{\dag})$, $\mathcal{F}_{22}=\mathcal{F}_{23}=\mathcal{F}_{33}
=S(\hat{r},\bm{\epsilon}_1\times\bm{\epsilon}_3^{\dag},\bm{\epsilon}_2\times\bm{\epsilon}_4^{\dag})$. In the numerical calculations, these operators $\mathcal{O}$ are replaced by several nonzero matrix elements by employing $\langle f|\mathcal{O}|i\rangle$, where the $|i\rangle$ and
$\langle f|$ stand for the spin-orbit wave functions for the initial and final states, respectively. For the $J^{PC}=0^{++}$ channel, $\langle f|\mathcal{E}_{12}|i\rangle\mapsto(-\sqrt{3},0)$, $\langle f|\mathcal{F}_{12}|i\rangle\mapsto(0,\sqrt{6})$, $\langle f|\mathcal{D}_{22}|i\rangle\mapsto\text{diag}(1,1)$, $\langle f|\mathcal{E}_{22}|i\rangle\mapsto\left(\begin{array}{cc}2&0\\0&-1\end{array}\right)$, and $\langle f|\mathcal{F}_{22}|i\rangle\mapsto\left(\begin{array}{cc}0&\sqrt{2}\\ \sqrt{2}&2\end{array}\right)$.

\section{The $X(3960)$ as the $D_s\bar{D}_s/D^*\bar{D}^*/D_s^*\bar{D}_s^*$ coupled resonance with $J^{PC}=0^{++}$}\label{sec3}

After prepared the OBE effective potentials, we would like to produce the scattering energy $\sqrt{s}$ dependence of the phase shifts $\delta(\sqrt{s})$ for the investigated coupled channel systems by varying the cutoff in the range from 0.80 GeV to 3.00 GeV. Here, the cutoff value $\Lambda$ in our OBE effective potentials is the only free parameter, it is related to the typical hadronic scale or the intrinsic size of hadrons. According to the experience of the nucleon-nucleon interactions \cite{Tornqvist:1993ng,Tornqvist:1993vu}, the reasonable values of the cutoff are taken around 1.00 GeV. These values are often adopted to the study of the interactions between the heavy hadrons.

With these obtained phase shifts, we can search for possible resonances, where the resonance generally emerges as the phase shifts satisfy $\delta(\sqrt{s_0})=(2n+1)\pi/2$ with $n=0,1,2,\ldots$. Here, the $s_0$ corresponds to the position of the obtained resonance, and its decay width can be estimated by $\Gamma=2/\left({d\delta(\sqrt{s})}/{{d}\sqrt{s}}\right)_{s={s_0}}$. Meanwhile, we also present the scattering energy $\sqrt{s}$ dependence of the scattering cross section $\sigma(\sqrt{s})=({4\pi}/{2m\sqrt{s}})\sum_{l=0}^{\infty}(2l+1)\text{sin}^2\delta_l(\sqrt{s})$. By these efforts, we can further check the resonant shapes.

\begin{figure}[!htbp]
\centering
\includegraphics[width=3.3in]{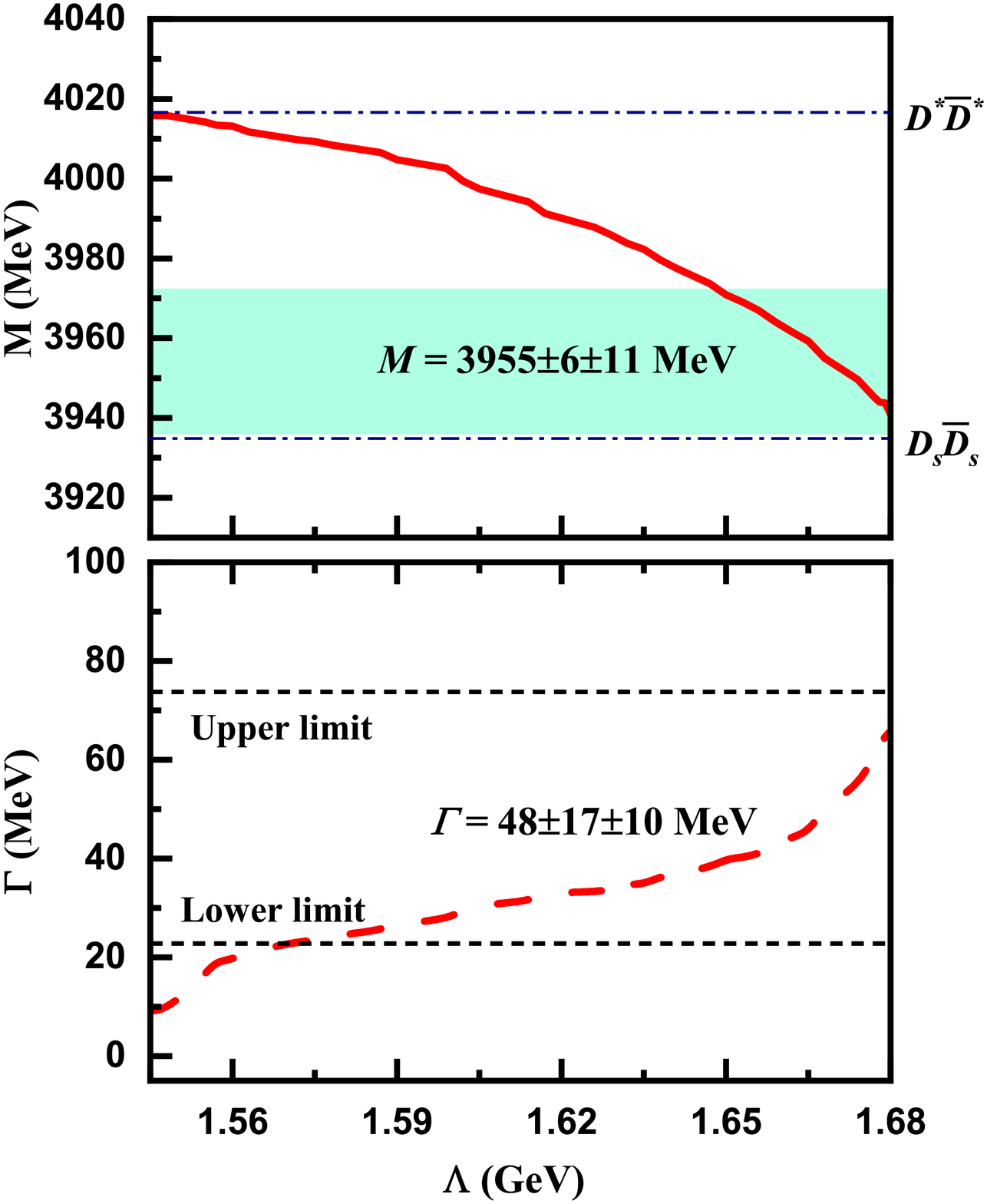}
\caption{The cutoff $\Lambda$ dependence of the obtained resonant mass $M$ for the isoscalar $D_s\bar{D}_s/D^*\bar{D}^*/D_s^*\bar{D}_s^*$ coupled systems with $J^{PC}=0^{++}$. Here, the shallow area corresponds to the reported experimental mass for the newly $X(3960)$ including the experimental uncertainty.}
\label{massX3960}
\end{figure}

When we produce the phase shifts for the isoscalar $D_s\bar{D}_s/D^*\bar{D}^*/D_s^*\bar{D}_s^*$ coupled systems with $J^{PC}=0^{++}$, we can obtain the resonance at the cutoff larger than 1.55 GeV. In Figures \ref{massX3960}, we present the cutoff $\Lambda$ dependence of the resonant mass $M$ and width $\Gamma$ for the isoscalar $D_s\bar{D}_s/D^*\bar{D}^*/D_s^*\bar{D}_s^*$ coupled systems with $J^{PC}=0^{++}$, respectively. Here, we find the resonance emerge at the cutoff $\Lambda=1.55$ GeV, which locates below the $D^*\bar{D}^*$ threshold. With the increasing of the cutoff value, the OBE effective potentials turn into stronger attractive, consequently, the resonant bind deeper and deeper. In particular, when the cutoff decreases to 1.65 GeV, the mass for this obtained resonance happens to overlap with the newly $X(3960)$ with experimental uncertainty. In addition, we identify the resonant width $\Gamma$ around 10 MeV at $\Lambda=1.55$ GeV. As the increasing of the cutoff value, the decay width becomes much larger. In the cutoff region from 1.57 GeV to 1.68 GeV, our results of the decay width varies from 21 MeV to 70 MeV, which is consistent with the experimental data of the newly $X(3960)$ with the experimental uncertainties.

\begin{figure}[!htbp]
\centering
\includegraphics[width=3.1in]{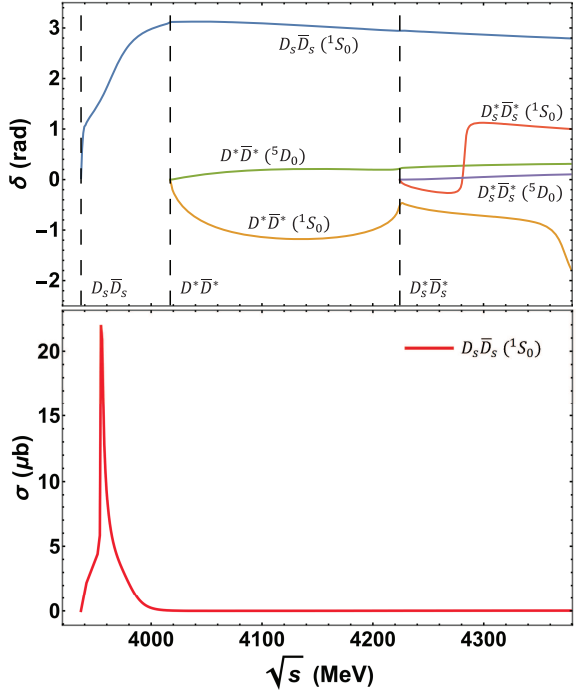}
\caption{The scattering energy $\sqrt{s}$ dependence of the phase shifts for all the investigated channels of the isoscalar $D_s\bar{D}_s/D^*\bar{D}^*/D_s^*\bar{D}_s^*$ coupled systems with $J^{PC}=0^{++}$ and the scattering cross section for the $D_s\bar{D}_s$ channel. Here, the cutoff is taken as $\Lambda=1.65$ GeV.}
\label{phase}
\end{figure}

The most important thing is that we can reproduce the mass and width for the newly observed $X(3960)$ in the cutoff region $\Lambda\geq1.65$ GeV, simultaneously. In Figure \ref{phase}, we present the scattering energy $\sqrt{s}$ dependence of the phase shifts for all the investigated channels of the isoscalar $D_s\bar{D}_s/D^*\bar{D}^*/D_s^*\bar{D}_s^*$ coupled systems with $J^{PC}=0^{++}$ and the scattering cross section for the $D_s\bar{D}_s$ channel at the cutoff $\Lambda=1.65$ GeV. Here, we can identify a resonance at the position $\sqrt{s}=3.97$ GeV as the phase shift of the $D_s\bar{D}_s({}^1S_0)$ channel crosses $\pi/2$. We can find a maximum cross section at the resonance energy, and the width is 38.13 MeV.

To summary, since the cutoff is close to the reasonable value \cite{Tornqvist:1993ng,Tornqvist:1993vu}, we can conclude that the newly $X(3960)$ can be explained as the isoscalar charmoniumlike resonance with $J^P=0^{++}$.

In this work, we further explore the roles of the $D^*\bar{D}^*$ and $D_s^*\bar{D}_s^*$ channels in generating the $X(3960)$ resonance. We produce the phase shifts for the $D_s\bar{D}_s/D^*\bar{D}^*$ coupled systems with $J^P=0^{++}$ and the $D_s\bar{D}_s/D_s^*\bar{D}_s^*$ coupled systems with $J^P=0^{++}$, respectively. Finally, our results indicate that there can exist resonant properties for these two coupled channel systems in the cutoff region $1.00<\Lambda<3.00$ GeV.

\begin{figure}[!htbp]
\centering
\includegraphics[width=3.3in]{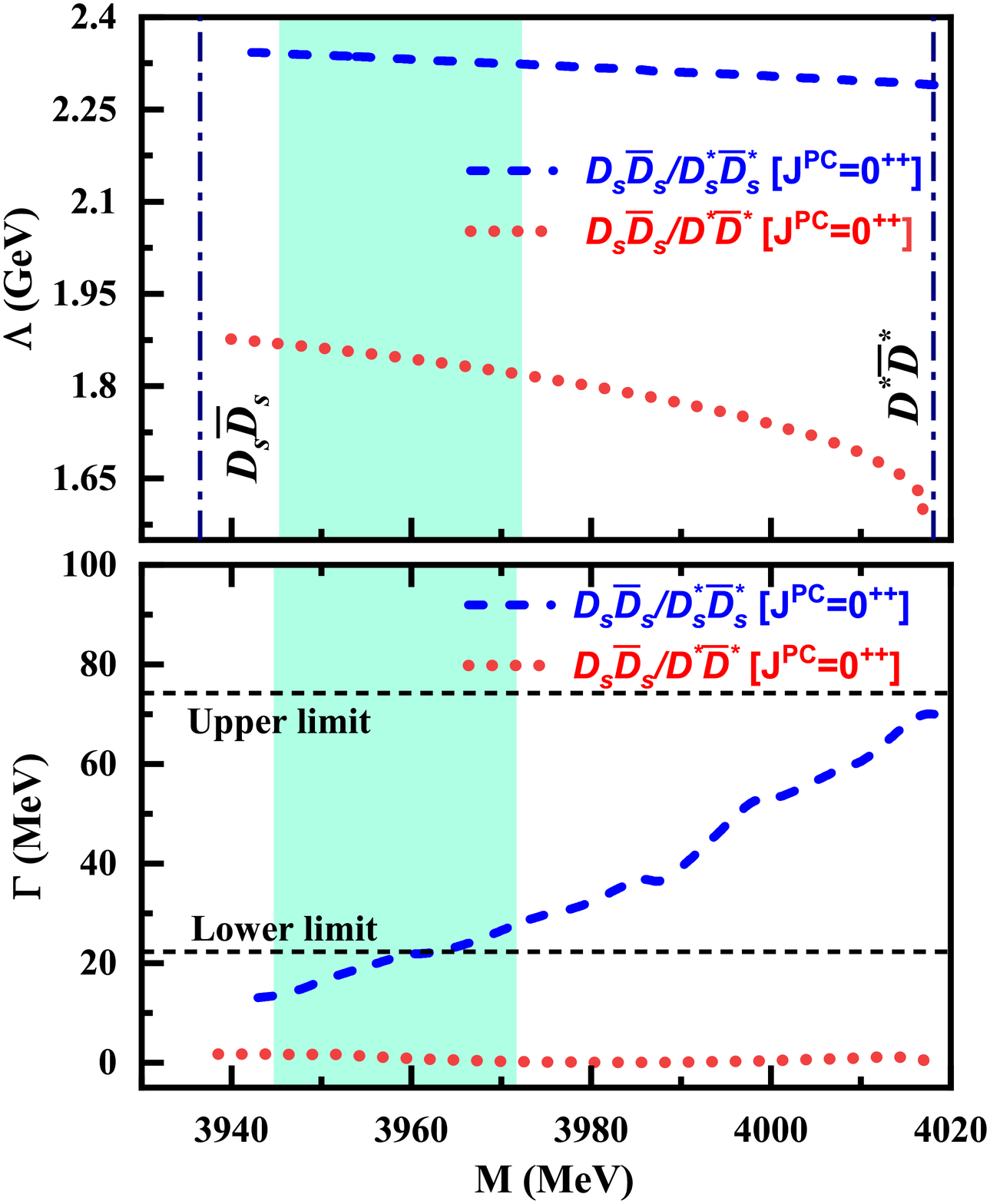}
\caption{The resonant mass dependence of the cutoff $\Lambda$ for the $D_s\bar{D}_s/D^*\bar{D}^*$ coupled systems with $J^P=0^{++}$ (red dotted line) and the $D_s\bar{D}_s/D_s^*\bar{D}_s^*$ coupled systems with $J^P=0^{++}$ (blue slash line). Here, the shallow area corresponds to the reported experimental mass for the newly $X(3960)$ including the experimental uncertainty. The short slash lines label the upper limit and lower limit for the width of the reported $X(3960)$.}
\label{fig2}
\end{figure}

In Figure \ref{fig2}, we present the obtained resonant mass dependence of cutoff value $\Lambda$ in the $D_s\bar{D}_s/D^*\bar{D}^*$ coupled systems with $J^P=0^{++}$ and the $D_s\bar{D}_s/D_s^*\bar{D}_s^*$ coupled systems with $J^P=0^{++}$, respectively. Here, we can see that for the $D_s\bar{D}_s/D^*\bar{D}^*$ coupled systems with $J^P=0^{++}$, the resonance appears at the cutoff $\Lambda$ larger than 1.60 GeV. In particular, when the cutoff increase to 1.85 GeV, the obtained resonant mass is 3957.03 MeV, which is close to the central mass of the $X(3960)$. Compared to the $D_s\bar{D}_s/D^*\bar{D}^*/D_s^*\bar{D}_s^*$ coupled systems with $J^{PC}=0^{++}$, the cutoff here is slightly larger, therefore, the OBE interactions in the $D_s\bar{D}_s/D^*\bar{D}^*$ coupled channel systems are a little weaker attractive than the $D_s\bar{D}_s/D^*\bar{D}^*/D_s^*\bar{D}_s^*$ interactions. Because the cutoff values still fall in the reasonable region, we can conclude that $D_s\bar{D}_s/D^*\bar{D}^*$ coupled channel systems provide strong enough attractive interactions to form a resonance, and the coupled channel effects originating from the $D_s^*\bar{D}_s^*$ systems play a minor and positive role.

For the $D_s\bar{D}_s/D_s^*\bar{D}_s^*$ coupled systems with $J^P=0^{++}$, the resonance emerges at the cutoff $\Lambda$ larger than 2.29 GeV. Obviously, this cutoff value $\Lambda$ is away from the cutoff value in the $D_s\bar{D}_s/D^*\bar{D}^*/D_s^*\bar{D}_s^*$ coupled systems with $J^{PC}=0^{++}$, which shows that the OBE interactions from the $D_s\bar{D}_s/D_s^*\bar{D}_s^*$ coupled channel systems are weaker attractive. Thus, the coupled channel effects originating from the $D^*\bar{D}^*$ system, which is discarded here, provide an important role in forming the $X(3960)$ as the resonance.

In the second subfigure in Figure \ref{fig2}, we also present the obtained resonant mass $M$ and width $\Gamma$ for the $D_s\bar{D}_s/D^*\bar{D}^*$ coupled systems with $J^P=0^{++}$ and the $D_s\bar{D}_s/D_s^*\bar{D}_s^*$ coupled systems with $J^P=0^{++}$, respectively. For the $D_s\bar{D}_s/D^*\bar{D}^*$ coupled systems with $J^P=0^{++}$, the obtained width is less than 1.00 MeV. It is too small compared to the experimental width of the newly $X(3960)$. Therefore, we can conclude that the $D_s^*\bar{D}_s^*$ channels play a very important role in generating the width of the $X(3960)$ as the charmoniumlike resonance.

However, for the $D_s\bar{D}_s/D_s^*\bar{D}_s^*$ coupled systems with $J^P=0^{++}$, the width varies from several MeV to seventy MeV in the mass region of $3940<M<4018$ MeV. When we align the resonant mass to the $X(3960)$, our theoretical results about the resonant width can fall into the experimental region for the $X(3960)$. In comparison with the results from the $D_s\bar{D}_s/D^*\bar{D}^*$ coupled systems with $J^P=0^{++}$, we find the $D_s^*\bar{D}_s^*$ channel affects the width for the $X(3960)$ a lot, which may be caused by the larger phase space for the $D_s^*\bar{D}_s^*$ decaying to the $D_sD_s$ final state. Anyway, if the newly $X(3960)$ can be regarded as the isoscalar charmoniumlike resonance, the contribution from the $D_s^*\bar{D}_s^*$ cannot be ignored.

From the current numerical results, both the $D^*\bar{D}^*$ and $D_s^*\bar{D}_s^*$ systems are very important in formation of the $X(3960)$ as the charmoniumlike resonance.

\section{Predictions of the $D^*\bar{D}^*/D_s\bar{D}_s^*/D_s^*\bar{D}_s^*$ coupled resonances with $J^{PC}=1^{+-}$}

In this section, we extend our study on the isoscalar $D^*\bar{D}^*/D_s\bar{D}_s^*/D_s^*\bar{D}_s^*$ interactions with $J^{PC}=1^{+-}$ by using the same model. After considering the $S-D$ wave mixing effects, the corresponding wave functions can be expanded as
\begin{eqnarray}
    |\Psi(1^{+-})\rangle &=& \psi_1(r) D^*\bar{D}^*|^3S_{1} \rangle + \psi_2(r) D^*\bar{D}^*|^3D_{1} \rangle  \nonumber \\
    & + & \psi_3(r)D^*\bar{D}^*|^5D_{1} \rangle  + \psi_4(r)D_s\bar{D}_s^*|^3S_{1} \rangle \nonumber \\
    & + & \psi_5(r)D_s\bar{D}_s^*|^3D_{1} \rangle + \psi_6(r) D_s^*\bar{D}_s^*|^3S_{1} \rangle \nonumber \\
    & + & \psi_7(r) D_s^*\bar{D}_s^*|^3D_{1} \rangle + \psi_8(r) D_s^*\bar{D}_s^*|^5D_{1} \rangle.
 \end{eqnarray}
Their OBE effective potentials are written as
\begin{eqnarray}
V &=& \left(\begin{array}{ccc}
    V^{D^*\bar{D}^*\to D^*\bar{D}^*}
        &V^{D_s\bar{D}_s^*\to D^*\bar{D}^*}      &V^{D_s^*\bar{D}_s^*\to D^*\bar{D}^*}\\
    V^{D^*\bar{D}^*\to D_s\bar{D}_s^*}
        &V^{D_s\bar{D}_s^*\to D_s\bar{D}_s^*}      &V^{D_s^*\bar{D}_s^*\to D_s\bar{D}_s^*}\\
    V^{D^*\bar{D}^*\to D_s^*\bar{D}_s^*}
        &V^{D_s\bar{D}_s^*\to D_s^*\bar{D}_s^*}      &V^{D_s^*\bar{D}_s^*\to D_s^*\bar{D}_s^*} \end{array}\right).\label{obe2}
\end{eqnarray}
Here, the subpotentials read as
\begin{eqnarray}
V^{D_s\bar{D}_s^*\to D_s\bar{D}_s^*} &=& -g_s^2\mathcal{Y}^{14}_{\Lambda,m_{\sigma}}
    -\frac{2}{9}\frac{g^2}{f_{\pi}^2}\mathcal{Z}^{41}_{\Lambda_0,m_{\eta0}}
    -\frac{\beta^2g_V^2}{2}\mathcal{Y}^{14}_{\Lambda,m_{\phi}}\nonumber\\
    &&+\frac{2}{3}\lambda^2g_V^2\mathcal{Z}^{\prime 41}_{\Lambda_0, m_{\phi0}},\label{eqa}\\
V^{D_s\bar{D}_s^*\to D^*\bar{D}^*} &=& -\frac{1}{3}\frac{g^2}{f_{\pi}^2}
    (\mathcal{Z}^{24}_{\Lambda_1,m_{K1}}+\mathcal{Z}^{42}_{\Lambda_1,m_{K1}})\nonumber\\
    &&+\frac{2}{3}\lambda^2g_V^2
    (\mathcal{X}^{24}_{\Lambda_1, m_{K^*1}}+\mathcal{X}^{42}_{\Lambda_1, m_{K^*1}}),\\
V^{D_s\bar{D}_s^*\to D_s^*\bar{D}_s^*} &=& -\frac{\sqrt{2}}{9}\frac{g^2}{f_{\pi}^2}
    (\mathcal{Z}^{34}_{\Lambda_2,m_{\eta2}}+\mathcal{Z}^{43}_{\Lambda_2,m_{\eta2}})\nonumber\\
    &&+\frac{\sqrt{2}}{3}\lambda^2g_V^2
    (\mathcal{X}^{34}_{\Lambda_2, m_{\phi2}}+\mathcal{X}^{43}_{\Lambda_2, m_{\phi2}}), \label{eqb}
\end{eqnarray}
with $\Lambda_i^2 =\Lambda^2-q_i^2$, $m_{ai}^2=m_{a}^2-q_i^2$, $i=0,1,2$, and
$q_0=m_{D_s^*}-m_{D_s}$, $q_1=(m_{D_s^*}^2-m_{D_s}^2)/4m_{D^*}$, $q_2=(m_{D_s^*}^2-m_{D_s}^2)/4m_{D_s^*}$. Here, we define
\begin{eqnarray*}
\mathcal{X}^{ij}_{\Lambda, m_a} &=&\left(-2\mathcal{E}_{ij}\nabla^2-(\mathcal{F}_{ij}^{\prime}-\mathcal{F}_{ij}^{\prime\prime})r\frac{\partial}{\partial
r}\frac{1}{r}\frac{\partial}{\partial r}\right)Y(\Lambda,m_a,r).
\end{eqnarray*}
In Table \ref{element}, we summary the corresponding matrix elements $\langle{}^{2s'+1}L'_{J'}|\mathcal{O}|{}^{2s+1}L_J\rangle$ for the spin-spin interactions and tensor force interactions operators in Eqs. (\ref{eqa})-(\ref{eqb}).

\renewcommand\tabcolsep{0.25cm}
\renewcommand{\arraystretch}{1.6}
\begin{table}[!hbtp]
\caption{Nonzero matrix elements $\langle{}^{2s'+1}L'_{J'}|\mathcal{O}|{}^{2s+1}L_J\rangle$ in various channels for the spin-spin interactions and tensor force interactions operators in Eqs. (\ref{eqa})-(\ref{eqb}). Here, $\mathcal{D}_{14}=\bm{\epsilon}_1\cdot\bm{\epsilon}_3^{\dag}$, $\mathcal{E}_{41}=\bm{\epsilon}_1\cdot\bm{\epsilon}_4^{\dag}$, $\mathcal{F}_{41}
=S(\hat{r},\bm{\epsilon}_1,\bm{\epsilon}_4^{\dag})$, $\mathcal{E}_{24}=\mathcal{E}_{34}=i\bm{\epsilon}_3^{\dag}\cdot(\bm{\epsilon}_4^{\dag}\times\bm{\epsilon}_2)$, $\mathcal{F}_{24}=\mathcal{F}_{34}= iS(\hat{r},\bm{\epsilon}_3^{\dag},\bm{\epsilon}_4^{\dag}\times\bm{\epsilon}_2)$, $\mathcal{F}_{24}^{\prime}=\mathcal{F}_{34}^{\prime}= iS(\hat{r},\bm{\epsilon}_4^{\dag},\bm{\epsilon}_2\times\bm{\epsilon}_3^{\dag})$, $\mathcal{F}_{24}^{\prime\prime}=\mathcal{F}_{34}^{\prime\prime}= iS(\hat{r},\bm{\epsilon}_2,\bm{\epsilon}_4^{\dag}\times\bm{\epsilon}_3^{\dag})$, $\mathcal{E}_{42}=\mathcal{E}_{43}=i\bm{\epsilon}_4^{\dag}\cdot(\bm{\epsilon}_3^{\dag}\times\bm{\epsilon}_1)$, $\mathcal{F}_{42}=\mathcal{F}_{43}= iS(\hat{r},\bm{\epsilon}_4^{\dag},\bm{\epsilon}_3^{\dag}\times\bm{\epsilon}_1)$, $\mathcal{F}_{42}^{\prime}=\mathcal{F}_{43}^{\prime}= iS(\hat{r},\bm{\epsilon}_3^{\dag},\bm{\epsilon}_1\times\bm{\epsilon}_4^{\dag})$, $\mathcal{F}_{42}^{\prime\prime}=\mathcal{F}_{43}^{\prime\prime}= iS(\hat{r},\bm{\epsilon}_1,\bm{\epsilon}_3^{\dag}\times\bm{\epsilon}_4^{\dag})$.} \label{element}
\begin{tabular}{c|c|c}
\toprule[1pt]\toprule[1pt]
$\mathcal{D}_{14}=\mathcal{E}_{41}$ &$\mathcal{F}_{41}$
   &$\mathcal{E}_{24}=\mathcal{E}_{34}=-\mathcal{E}_{42}=-\mathcal{E}_{43}$
\\
 $\left(\begin{array}{cc}1  &0\\ 0 &1\end{array}\right)$
    &$\left(\begin{array}{cc}0  &-\sqrt{2}\\ -\sqrt{2}  &1\end{array}\right)$
    &$\left(\begin{array}{cc}\sqrt{2} &0\\ 0  &\sqrt{2}\\  0 &0\end{array}\right)$\\\hline
 $\mathcal{F}_{24}=\mathcal{F}_{34}$
    &$\mathcal{F}_{24}^{\prime}=\mathcal{F}_{34}^{\prime}$
    &$\mathcal{F}_{24}^{\prime\prime}=\mathcal{F}_{34}^{\prime\prime}$\\
 $-\mathcal{F}_{42}^{\prime}=-\mathcal{F}_{43}^{\prime}$
    &$-\mathcal{F}_{42}=-\mathcal{F}_{43}$
    &$-\mathcal{F}_{42}^{\prime\prime}=-\mathcal{F}_{43}^{\prime\prime}$\\
 $\left(\begin{array}{cc}0 &1\\ 1 &-\frac{1}{\sqrt{2}}\\ \sqrt{3} &\sqrt{\frac{3}{2}}\end{array}\right)$
  &$\left(\begin{array}{cc}0 &1\\ 1 &-\frac{1}{\sqrt{2}}\\ -\sqrt{3} &-\sqrt{\frac{3}{2}}\end{array}\right)$
  &$\left(\begin{array}{cc} 0&2\\ 2&-\sqrt{2}\\0&0\end{array}\right)$\\
 \bottomrule[1pt]
\bottomrule[1pt]
\end{tabular}
\end{table}

\begin{figure}[!htbp]
\centering
\includegraphics[width=3.3in]{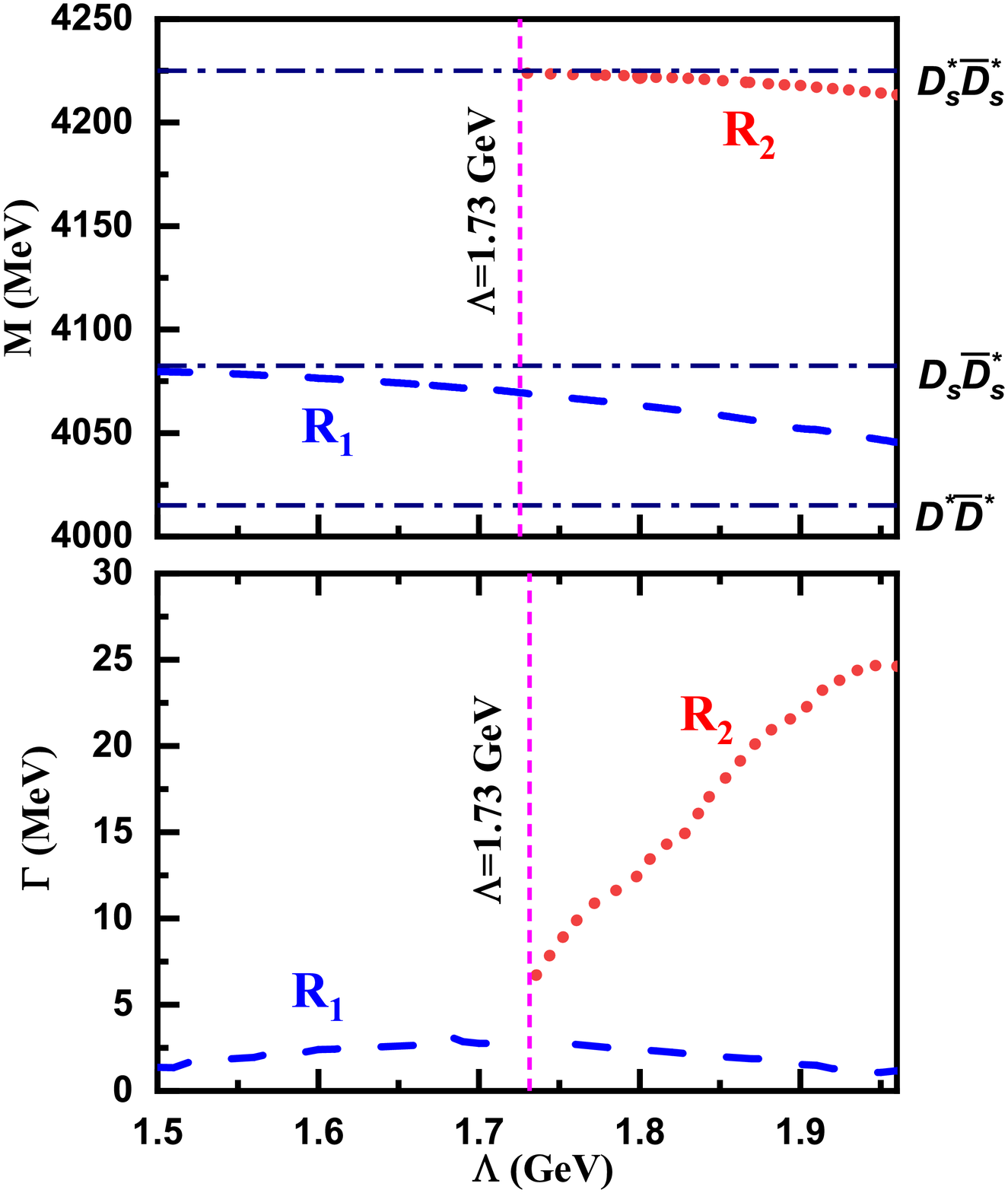}
\caption{The cutoff $\Lambda$ dependence of the resonant parameters (mass $M$ and decay width $\Gamma$) for the isoscalar $D^*\bar{D}^*/D_s\bar{D}_s^*/D^*\bar{D}^*$ interactions with $J^{PC}=1^{+-}$.}
\label{X1}
\end{figure}

After that, we produce the phase shifts for the isoscalar $D^*\bar{D}^*/D_s\bar{D}_s^*/D_s^*\bar{D}_s^*$ interactions with $J^{PC}=1^{+-}$. As shown in Figure \ref{X1}, we present the cutoff $\Lambda$ dependence of the resonant parameters (mass $M$ and decay width $\Gamma$) for the isoscalar $D^*\bar{D}^*/D_s\bar{D}_s^*/D_s^*\bar{D}_s^*$ coupled systems with $J^{PC}=1^{+-}$. When we tune the cutoff at $\Lambda=$1.50 GeV, we can obtain a possible charmoniumlike resonant candidate (labeled as $R_1$) with mass $M=4079.58$ MeV, which just locates below the $D_s\bar{D}_s^*$ threshold $M_{D_s\bar{D}_s^*}=4080.54$ MeV. The width is 1.37 MeV. When the cutoff reaches at $\Lambda=$1.72 GeV, the attractions of the OBE effective potentials become stronger, the mass and decay width for the $R_1$ state turn into 4069.57 MeV and 2.80 MeV, respectively. In addition, we find another possible charmoniumlike resonant candidate (labeled as $R_2$) slightly located below the $D_s^*\bar{D}_s^*$ threshold, whose width is 5.93 MeV. As shown in Figure \ref{X1}, the width for the $R_2$ resonance is larger than that for the $R_1$ resonance. Especially, in the cutoff region $1.70\leq\Lambda\leq2.00$ GeV, the width for the $R_2$ resonance can reach around 20 MeV. And the decay width for the $R_1$ resonance is still less than 3.00 MeV.

\begin{figure}[!htbp]
\centering
\includegraphics[width=3.1in]{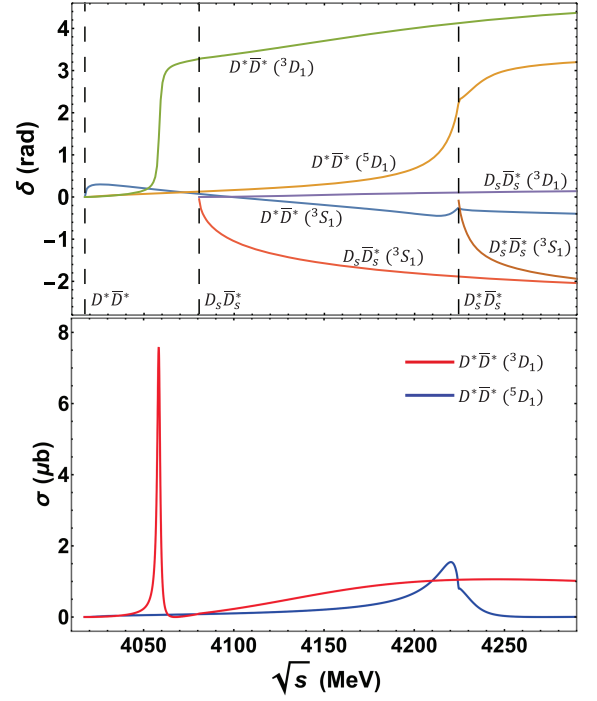}
\caption{The scattering energy $\sqrt{s}$ dependence of the phase shifts for all the investigated channels of the $D^*\bar{D}^*/D_s\bar{D}_s^*/D_s^*\bar{D}_s^*$ coupled resonances with $J^{PC}=1^{+-}$ and the scattering cross section for the $D^*\bar{D}^*$ channel. Here, the cutoff is taken as $\Lambda=1.65$ GeV.}
\label{phase1}
\end{figure}

In Figure \ref{phase1}, we present the scattering energy $\sqrt{s}$ dependence of the phase shifts for all the investigated channels of the $D^*\bar{D}^*/D_s\bar{D}_s^*/D_s^*\bar{D}_s^*$ coupled resonances with $J^{PC}=1^{+-}$ and the scattering cross section for the $D^*\bar{D}^*$ channel. We can identify two 

Here, we also find that the cutoff values are close to those in the case of the newly observed $X(3960)$ as the isoscalar charmoniumlike resonance with $J^P=0^{++}$ as shown in Figure \ref{massX3960}. Therefore, if the $X(3960)$ can be assigned as a charmoniumlike resonance, the $R_1$ and $R_2$ can be also possible charmoniumlike resonant candidates. Their masses are very close to the $D_s\bar{D}_s^*$ and $D_s^*\bar{D}_s^*$ thresholds, respectively.

The close threshold properties remind us the predictions in our previous paper \cite{Wang:2021aql}, when we systematically study the interactions between a charmed (charmed-strange) meson and an anti-charmed (anti-charm-strange) meson by using the OBE model, we find the $D_s\bar{D}_s^*$ state with $J^{PC}=1^{+-}$ and the $D_s^*\bar{D}_s^*$ state with $J^{PC}=1^{+-}$ can be possible molecular candidates. And the coupled channel effects play an important role in generating the $D_s\bar{D}_s^*$ molecular state with $J^{PC}=1^{+-}$. In comparison with the previous results, we can further conclude that these two possible charmoniumlike resonance obtained in this paper are not new structures but the $D_s\bar{D}_s^*$ molecule with $J^{PC}=1^{+-}$ and the $D_s^*\bar{D}_s^*$ molecule with $J^{PC}=1^{+-}$.

\section{Summary}\label{sec4}

Very recent, the LHCb Collaboration observed a neutral state $X(3960)$ in the $D_s^+ D_s^-$ invariant mass spectrum of the $B^+ \to D_s^+ D_s^- K^+$ process \cite{LHCb:Tcc}. According to the mass and decay properties, the $X(3960)$ is very likely to be a charmoniumlike exotic state. Up to now, the inner structure for the newly $X(3960)$ is still open to discuss. In this work, we propose the $X(3960)$ as the isoscalar $\mathcal{D}\overline{\mathcal{D}}-$type charmoniumlike resonance with $J^P=0^{++}$, where $\mathcal{D}$ stands for the $S-$wave charmed and charmed-strange mesons.

In order to examine our proposal, we analyze the phase shifts for the $D_s\bar{D}_s/D^*\bar{D}^*/D_s^*\bar{D}_s^*$ coupled systems with $J^{PC}=0^{++}$ after adopting the OBE effective potentials and considering the $S-D$ wave mixing effects. Our results show that there can exist a possible charmoniumike resonance in the reasonable cutoff input. The obtained resonant mass and width are consistent with the experimental data of the newly $X(3960)$. Here, we also find that both the $D^*\bar{D}^*$ and $D_s^*\bar{D}_s^*$ channels will play important roles in binding and impacting on the decay width of the $X(3960)$ as the charmoniumlike resonance, respectively.

In addition, we adopt the same OBE model to study the isoscalar $D^*\bar{D}^*/D_s\bar{D}_s^*/D_s^*\bar{D}_s^*$ interactions with $J^{PC}=1^{+-}$. Finally, we obtain two possible charmoniumlike structures, which can correspond to the $D_s\bar{D}_s^*$ molecular state with $J^{PC}=1^{+-}$ and the $D_s^*\bar{D}_s^*$ molecular state with $J^{PC}=1^{+-}$ \cite{Wang:2021aql}, whose widths are in orders of magnitudes of several and several to several tens MeV, respectively. The $\eta_c\phi$ and $J/\psi\eta^{(\prime)}$ can be the important two-body hidden-charm decay channels for these two bound states. The $D_s\bar{D}_s^*$ channel is also the only open-charm decay mode for the $D_s^*\bar{D}_s^*$ molecular state with $J^{PC}=1^{+-}$.

The experimental progress, especially the improvement of experimental techniques and the accumulation of the experimental data, will provide us a good chance to explore the underlying mechanism or inner structures for the new exotic states. We look forward to the further experiment to verify our proposal.

\section*{ACKNOWLEDGMENTS}

This project is supported by the National Natural Science Foundation of China under Grants No. 12305139 and No. 12305087. Rui Chen is supported by the Xiaoxiang Scholars Programme of Hunan Normal University. Qi Huang is supported by the Start-up Funds of Nanjing Normal University under Grant No. 184080H201B20.

\end{document}